\documentclass{article}
\usepackage{authblk}
\usepackage{graphicx}
\usepackage{a4wide}
\usepackage[pdftex]{hyperref}

\title{Suppression of the optical crosstalk in a multi-channel silicon photomultiplier array}

\author[1]{Takahiko~Masuda}
\author[2]{Daniel~G.~Ang}
\author[3]{Nicholas~R.~Hutzler}
\author[2]{Cole~Meisenhelder}
\author[1]{Noboru~Sasao}
\author[1]{Satoshi~Uetake}
\author[2,4]{Xing~Wu}
\author[4]{David~DeMille}
\author[5]{Gerald~Gabrielse}
\author[1,2]{John~M.~Doyle}
\author[1]{Koji~Yoshimura}

\affil[1]{Research Institute for Interdisciplinary Science, Okayama University, Okayama 700-8530, Japan}
\affil[2]{Department of Physics, Harvard University, Cambridge, Massachusetts 02138, USA}
\affil[3]{Division of Physics, Mathematics, and Astronomy, California Institute of Technology, Pasadena, California 91125, USA}
\affil[4]{James Franck Institute and Department of Physics, University of Chicago, Chicago, Illinois 60637, USA}
\affil[5]{Center for Fundamental Physics, Northwestern University, Evanston, Illinois 60208, USA}

\begin{document}
\maketitle


\begin{abstract}
We propose and study a method of optical crosstalk suppression for silicon photomultipliers (SiPMs) using optical filters.
We demonstrate that attaching absorptive visible bandpass filters to the SiPM can substantially reduce the optical crosstalk.
Measurements suggest that the absorption of near infrared light is important to achieve this suppression. 
The proposed technique can be easily applied to suppress the optical crosstalk in SiPMs in cases where filtering near infrared light is compatible with the application.
\end{abstract}


\section{Introduction} \label{sec:intro}

A silicon photomultiplier (SiPM) consists of an array of small avalanche photodiode pixels that are built on a silicon substrate. These pixels are operated in Geiger Mode and are sensitive to single photon injection, where a single photon stimulates an avalanche process in a single pixel that discharges the pixel completely. All pixels are electrically connected in parallel, so the output signal is an analog pulse whose charge is proportional to the total number of fired pixels.
The advantage of SiPM is the higher quantum efficiency compared to photomultiplier tube (PMT), and lower dark current compared to avalanche photodiode (APD), which makes SiPM ideal for low-photon flux situations.
Among the myriad of photon detectors sensitive to light ranging from the near ultraviolet to near infrared (NIR) region, SiPMs are rapidly expanding in their applicability in many fields \cite{Acerbi2019,Yamamoto2019} from high energy physics \cite{Garutti2011}, astro-particle physics \cite{Anderhub2013}, fluorescence detection \cite{Song2007}, Raman spectroscopy \cite{Zhang2010}, and biology-related measurements \cite{Mora2015,Modi2019}, to applications including light detection and ranging \cite{Agishev2013}, and medical applications \cite{Otte2005}.

SiPMs are especially ideal photo detectors when the photon number resolution is important. That is because the output pulse triggered by a single photon is quite uniform, with low shape variability and total output pulse charge proportional to the number of incident photons.
In our current application, we are developing a photon detector based on a SiPM for the ACME experiment \cite{Baron2014,Andreev2018}. The ACME experiment  aims to observe a non-zero permanent electric dipole moment (EDM) of the electron as concrete evidence of new physics beyond the Standard Model of particle physics. 
The fundamental source of the signal in ACME is laser-induced fluorescence (LIF) detection of thorium monoxide (ThO) molecules. ThO molecules in the $H^3\Delta_1$ electronic state \cite{Vutha2011} are excited to the \textit{I} state by applying 703\,nm laser light. The \textit{I} state spontaneously decays to the ground state with 512\,nm photon emission \cite{Kokkin2015}. The accurate determination of the number of 512\,nm photons emitted in different experimental configurations determines the sensitivity of the EDM measurement.

Optical crosstalk (OCT) in a SiPM is a factor limiting the photon number resolution. 
OCT is caused by secondary photons generated and emitted during the avalanche process triggered by the first incident photon. If one of the secondary photons is detected by one of the surrounding pixels, the pixel is then fired \cite{Rech2008}.
Because the OCT process generates an additional signal pulse associated with the original signal pulse, OCT acts as a gain variation, degrading the photon number resolution; it is known as excess noise \cite{VanVliet1979}.
Besides the photon number resolution, OCT significantly reduces the dynamic range of the downstream readout electronics because it generates signals with two or more times larger pulse heights than expected for a single photon.
OCT is also a problem, for example, in very-high-energy ground-based gamma-ray telescopes which employ SiPMs as photo sensors because OCT can degrade the energy resolution and increase the accidental trigger rate \cite{Anderhub2013,Acharya2013}.
OCT is one of the characteristics that has always been a concern in the use of SiPMs; therefore, OCT suppression technology is an active area of research \cite{Sul2013}.

In this study, we focus on an approach to reduce OCT using the fact that the OCT is affected by wavelength-selective optical filters, such as longpass filters (LPFs) or bandpass filters (BPFs).
 In any application where the objective light wavelength is fixed or narrow, a wavelength-selective optical filter is a potentially useful tool to reduce spurious photon contributions whose wavelength is different from the objective photons so as to improve the signal-to-noise ratio (SNR). In atomic spectroscopy or similar measurements, optical filters are commonly used in tandem with photo detectors. 
 For the ACME experiment mentioned above, we intend to place optical filters in front of SiPMs in order to eliminate scattered 703\,nm photons that originate from the probe laser used during the LIF measurement along with other residual light from the experimental environment.
But along with the background rejection effect, optical filters also affect the SiPM detector performance in multiple ways \cite{Mazzillo2017a, Mazzillo2017b, Mazzillo2018a, Mazzillo2018b}. These phenomena can be explained in terms of the detailed behavior of OCT.

In this paper, we present a comprehensive study of the OCT in a SiPM system with optical filters and propose a method that suppresses OCT by gluing absorptive filters onto the SiPM surface. Section~\ref{sec:measurement} presents the measurement of the OCT probabilities with various optical filters. Discussion of the results is presented in Section~\ref{sec:discussion}. Section~\ref{sec:conclusion} presents a summary and conclusions of the study.

\section{Optical crosstalk measurement\label{sec:measurement}}

\subsection{Measurement setup and method}

 The SiPM array we used in this study is a commercial surface mount 16-channel SiPM array S13361-6075NE-04 (Hamamatsu Photonics K.K.). The 16 channels are aligned in a $4 \times 4$ matrix. Each channel has a sensitive area of $6 \times 6\, \mathrm{mm^2}$ that consists of 6400 pixels whose size is $75 \times 75\,\mathrm{\mu m^2}$. The geometrical fill factor is 82\%. The gap between channels is $200\, \mathrm{\mu m}$ and the outer dimension of the array is $25 \times 25\, \mathrm{mm^2}$. The array is sealed by transparent epoxy resin with a refractive index of about 1.55 as a surface protection window.
 
\begin{figure}[h!]
\centering
\includegraphics[width=12cm, bb=0 0 338 172]{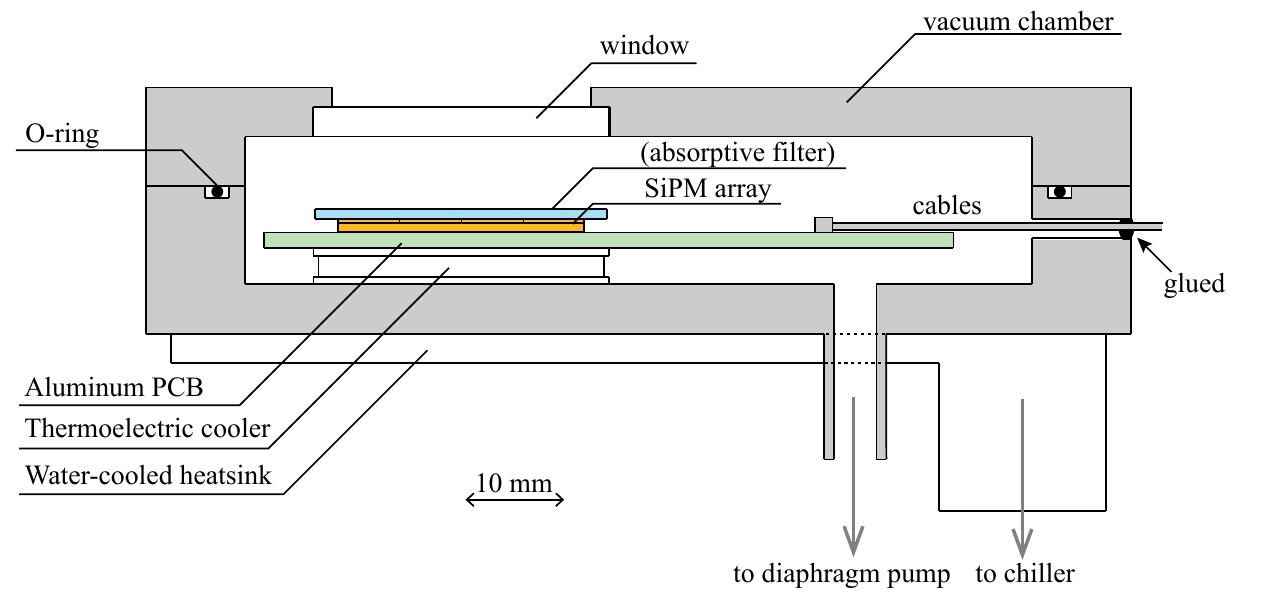}
\caption{Cutout view of the SiPM module.}
\label{fig:modulecutout}
\end{figure}

\begin{figure}[h!]
\centering
\includegraphics[width=7cm, bb=0 0 2375 1742]{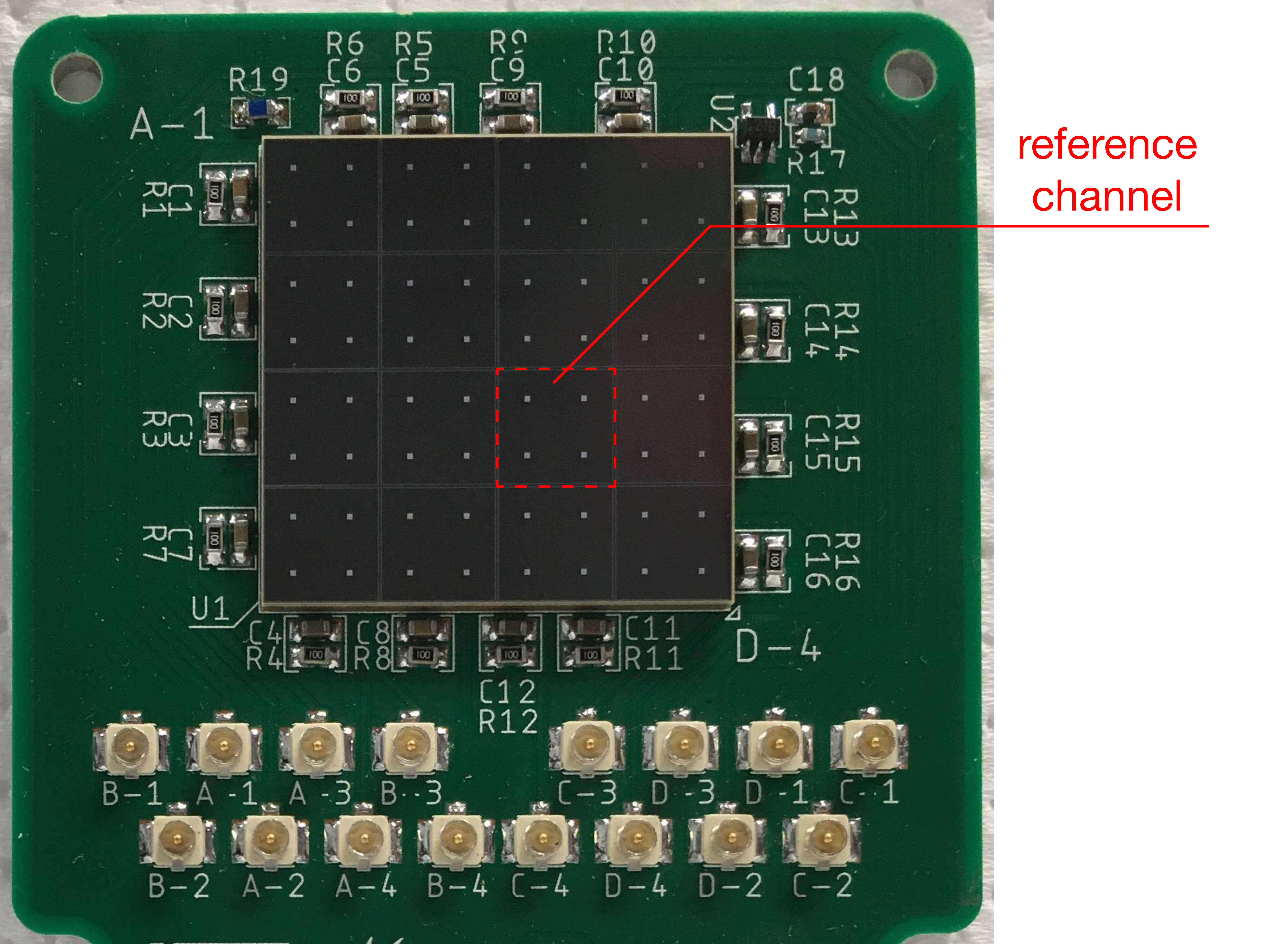}
\caption{Photograph of the SiPM module. The channel surrounded in the red square is the reference channel.}
\label{fig:sipmpicture}
\end{figure}

 The SiPM array was housed in a vacuum chamber in which the vacuum level was $\sim$100\,Pa. Fig.~\ref{fig:modulecutout} shows the schematic of the vacuum chamber and Fig.~\ref{fig:sipmpicture} shows the photograph of the SiPM mounted on a printed circuit board (PCB). We mounted the SiPM array onto a PCB made of a 1.5-mm-thick aluminum plate to cool the SiPM from the backside of the PCB. A thermoelectric cooler element (Thorlabs, Inc. TECH4) was attached between the PCB and the chamber wall using grease (Apiezon N). A platinum temperature sensor was mounted on the PCB and the PCB temperature was stabilized at $-10\,^\circ \mathrm{C}$ to reduce the dark count rate. The heat load was transferred to a water-cooled heatsink on the backside of the chamber. The temperature of the heatsink was stabilized by a thermoelectric chiller (ThermoTek T257P-20).
During the study, the top vacuum window was covered by a black sheet to prevent stray lights from outside.
 
\begin{figure}[h!]
\centering
\includegraphics[width=13cm, bb=0 0 791 171]{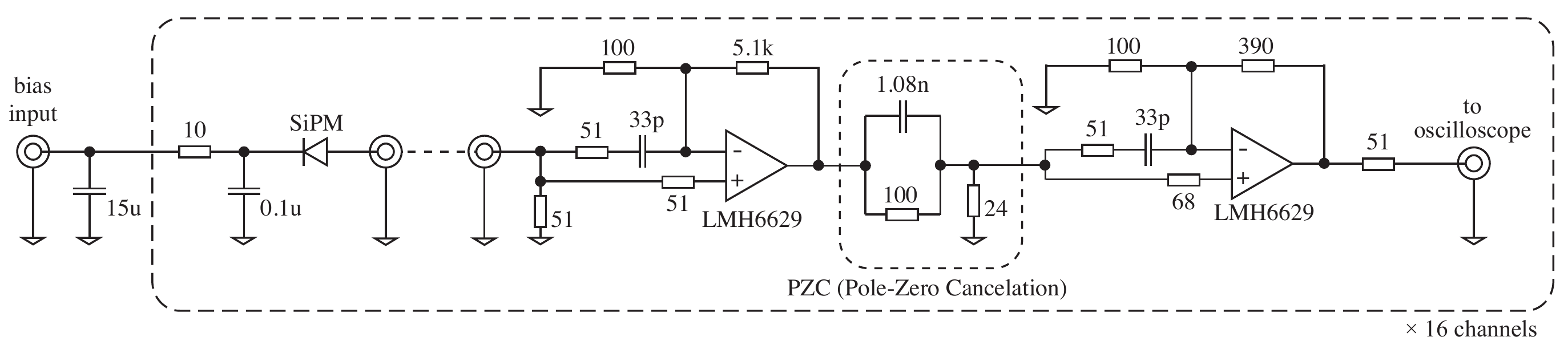}
\caption{Circuit diagram for a channel.}
\label{fig:circuit}
\end{figure}

The bias voltage, signal lines from the array, and the temperature sensor were connected via 50\,$\Omega$ small coaxial cables (Hirose U.FL) to the outside of the chamber. All 16 channels were connected in parallel and were operated at the same bias voltage so that the overvoltage was 3.0\,V. The output signal from each channel was amplified by a preamplifier. We designed a dedicated pole-zero cancellation circuit \cite{Gola2013} to shorten the pulse and improve the timing resolution; the resultant pulse duration had a full width at half maximum (FWHM) of $\sim$25\,ns. The circuit diagram of one channel is shown in Fig.~\ref{fig:circuit}.

The output signal was recorded by a 4-channel oscilloscope (NI PXIe-5162) with a sampling rate of 1.25\,GHz and a -3\,dB bandwidth of 175\,MHz.

 Because we used a multichannel SiPM, we considered both intra-channel OCT and inter-channel OCT. The intra-channel OCT is a phenomenon occurring in the same channel; one of the secondary photons is detected in another pixel in the same channel and the output pulse has a 2-photoelectron (p.e.) pulse height. In contrast, the inter-channel OCT is a crosstalk between two of the channels: a secondary photon propagates to another channel through the surface protection window and generates an additional p.e. pulse at the channel; therefore, the two channels output single p.e. pulses at the same time. Inter-channel OCT has been recently reported and studied by a ray-tracing simulation \cite{Nakamura2019}. 
 
 We focused our analysis on the OCT from a particular central single channel, which hereafter we refer to the channel as the reference channel, as indicated in Fig.~\ref{fig:sipmpicture}. The digitizer recorded output pulses from the reference channel using a leading edge timing trigger.
As a source of primary pulses, we used dark counts to simulate photon injection.

\begin{figure}[h!]
\centering
\includegraphics[width=7cm, bb=0 0 513 356, clip]{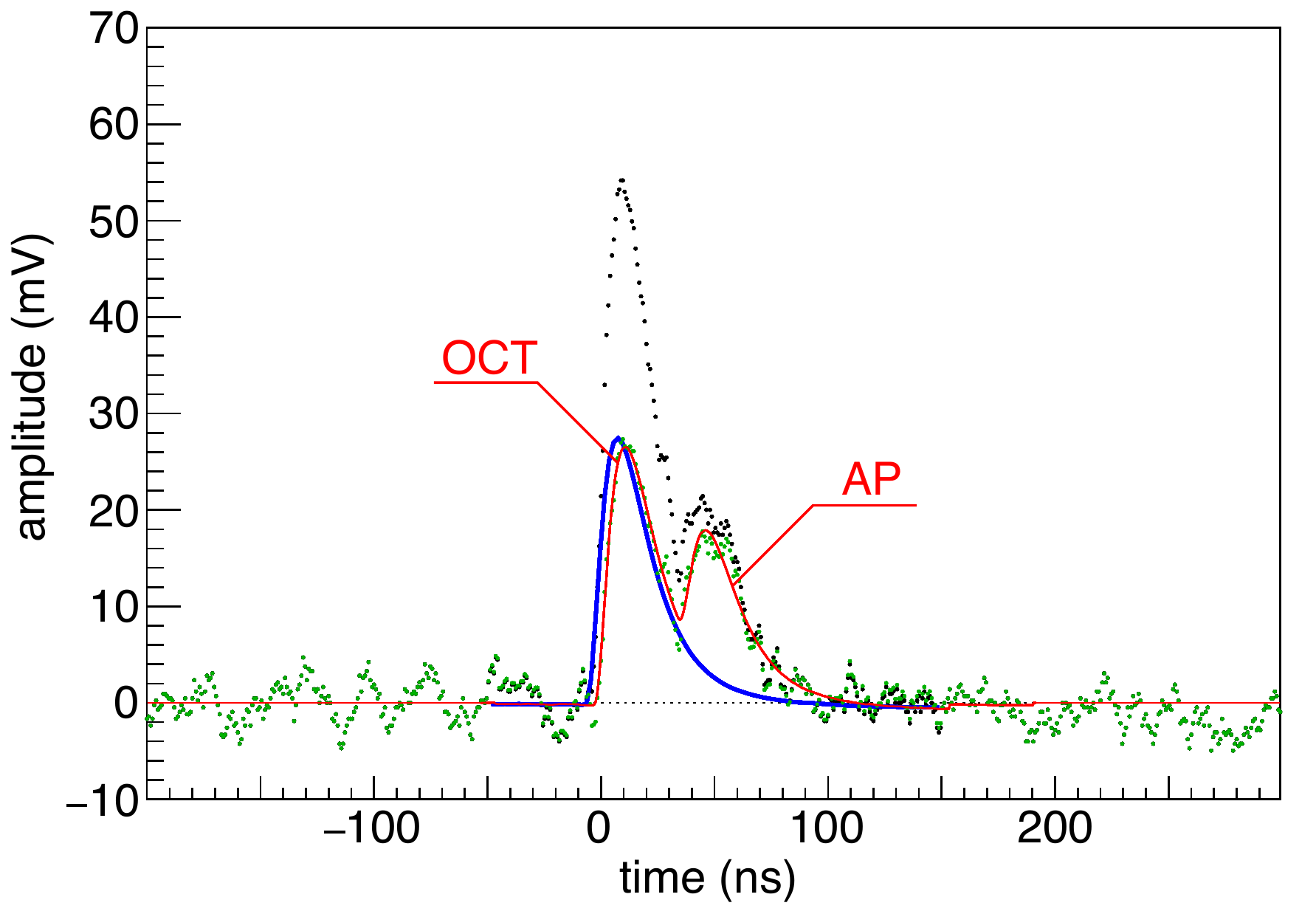}
\caption{Example of the template pulse fitting for both OCT and AP existed. The black points represent the raw temporal trace of the reference channel. The blue line is the pulse template. The green points show the subtracted temporal trace and the red curve shows the template fitting result based on the green points. The origin of the horizontal axis is the trigger timing at the leading edge.}
\label{fig:templatefit}
\end{figure}

\begin{figure}[h!]
\centering
\includegraphics[width=13cm, bb=0 0 2308 806, clip]{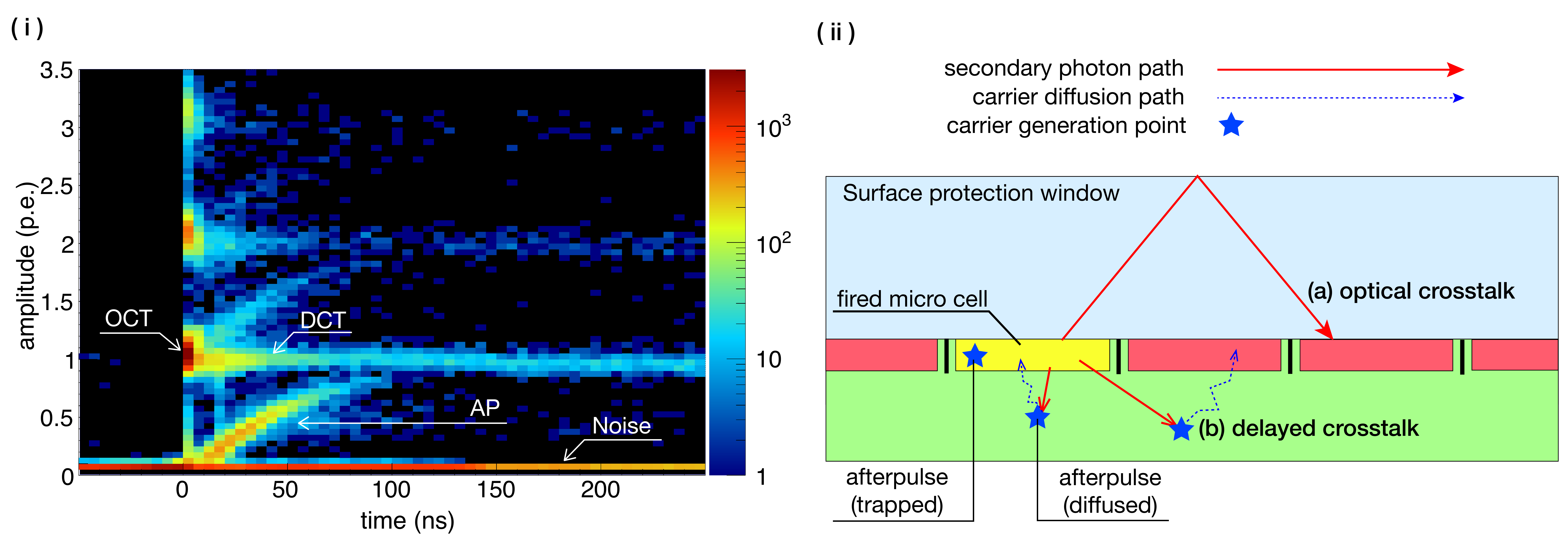}
\caption{(i) Amplitude-delay two-dimensional histogram of the associated pulses obtained by the template pulse fitting. The vertical axis is normalized to single p.e. amplitude. OCT: optical crosstalk; DCT: delayed crosstalk; AP: afterpulse.
(ii) schematic view of the optical crosstalk and the delayed crosstalk. The path labeled (a) is the optical crosstalk in which a secondary photon propagates to another micro cell. The path labeled (b) shows the delayed crosstalk mechanism: the secondary photon going to the backside is absorbed in silicon and generates a carrier, then the carrier moves to another micro cell with a certain probability due to diffusion. Two after pulse mechanisms are also shown: after pulse (diffused) is a similar process as the DCT but the diffused carrier returns to the original cell, and after pulse (trapped) is caused by a carrier trapped in a crystal defect in the fired micro cell \cite{Rosado2015}.}
\label{fig:dctmechanism}
\end{figure}

 To evaluate the intra-channel OCT probability, we analyzed the output traces with a template pulse fitting as shown in Fig.~\ref{fig:templatefit}. The pulse template was obtained by accumulating single p.e. pulses and averaging them in advance. We first subtracted one template pulse from the raw time trace so that the remaining component could only contain associated pulses, if they exist. The template fitting was able to distinguish multiple pulses in the remaining component with a timing resolution of $\sim$2.0\,ns.
 An example of the amplitude-delay distribution of the associated pulses is shown in Fig.~\ref{fig:dctmechanism}(i). In addition to the OCT, which is the peak component located at $t=0$, afterpulse (AP) and delayed crosstalk (DCT) components were clearly distinguished. The AP is the component that distributes below 1 p.e. amplitude and is rising as $(1-\exp)$, which occurs in the same cell \cite{Rosado2015}. The DCT is the component whose amplitude is almost 1 p.e. and timing is delayed, which is caused by secondary photons going to the backside and the carrier diffusion process \cite{Nagy2014} as shown in Fig.~\ref{fig:dctmechanism}(ii). We set a region of interest to distinguish OCT components from other components as 0--6\,ns along the horizontal axis and amplitude of more than 0.8 p.e. along the vertical axis in this study.
 
On the other hand, the evaluation of the inter-channel OCT probability is much simpler because neither AP nor DCT occurs and any single p.e. pulses at the same time between channels can be considered as OCT. We acquired time traces of other channels simultaneously, and calculated the probability that the other channel has single p.e. pulses at the same time as the reference channel. 
Since we cannot distinguish which of the two channels was the source of OCT, we assumed the OCT probabilities were the same in both directions; hence,  the OCT probability was approximately the half of the probability that the other channel has single p.e. pulses.

\subsection{OCT measurement with visible bandpass filters}
\begin{figure}[h]
\centering
\includegraphics[width=13cm, bb=0 0 1532 625, clip]{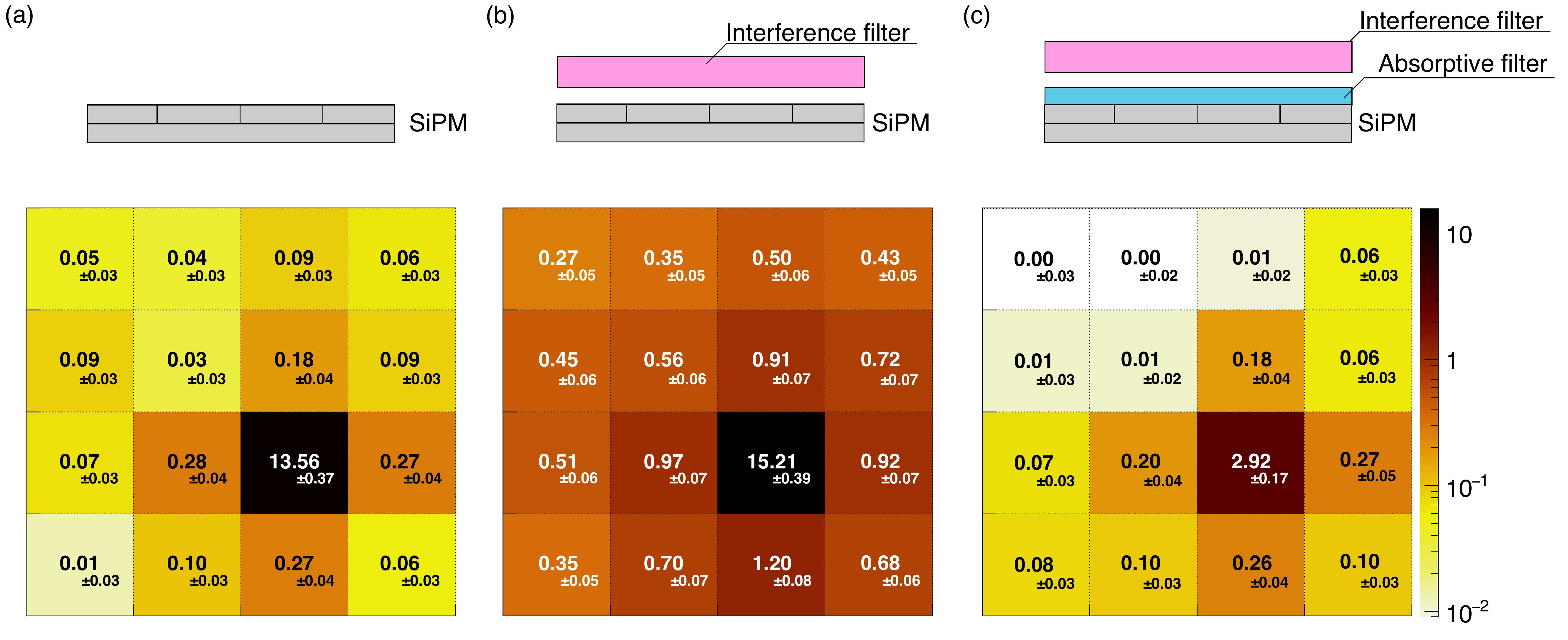}
\caption{OCT distributions with bandpass filters. (a) No filter. (b) interference BPF is put on the SiPM. (c) Absorptive BPF is glued on the SiPM and the interference BPF is put on it. The number in each channel represents the OCT probability with the statistical uncertainty from the reference channel in unit of \%.}
\label{fig:bpfeffect}
\end{figure}

 When a good extinction ratio between the transmission band and the blocking band is needed, an interference BPF is usually used.
We used the Semrock FF01-520/70 as a green interference BPF whose center wavelength is 520\,nm and full width at half maximum is 70\,nm. The average transmittance in the transmission band is more than 93\% and the optical density in the blocking bands is more than 5. We put the interference BPF above the SiPM surface with a clearance of $\sim$1\,mm. 
Figures~\ref{fig:bpfeffect}(a) and \ref{fig:bpfeffect}(b) show the OCT probabilities without and with the interference BPF, respectively. The distribution in the array clearly shows that the OCT, namely secondary photons, were spreading from the reference channel to the other channels. The total OCT probability in the 16-channels increased by a factor of 1.6 (15.3\% to 24.7\%).

\begin{figure}[h]
\centering
\includegraphics[width=13cm, bb=0 0 2269 911, clip]{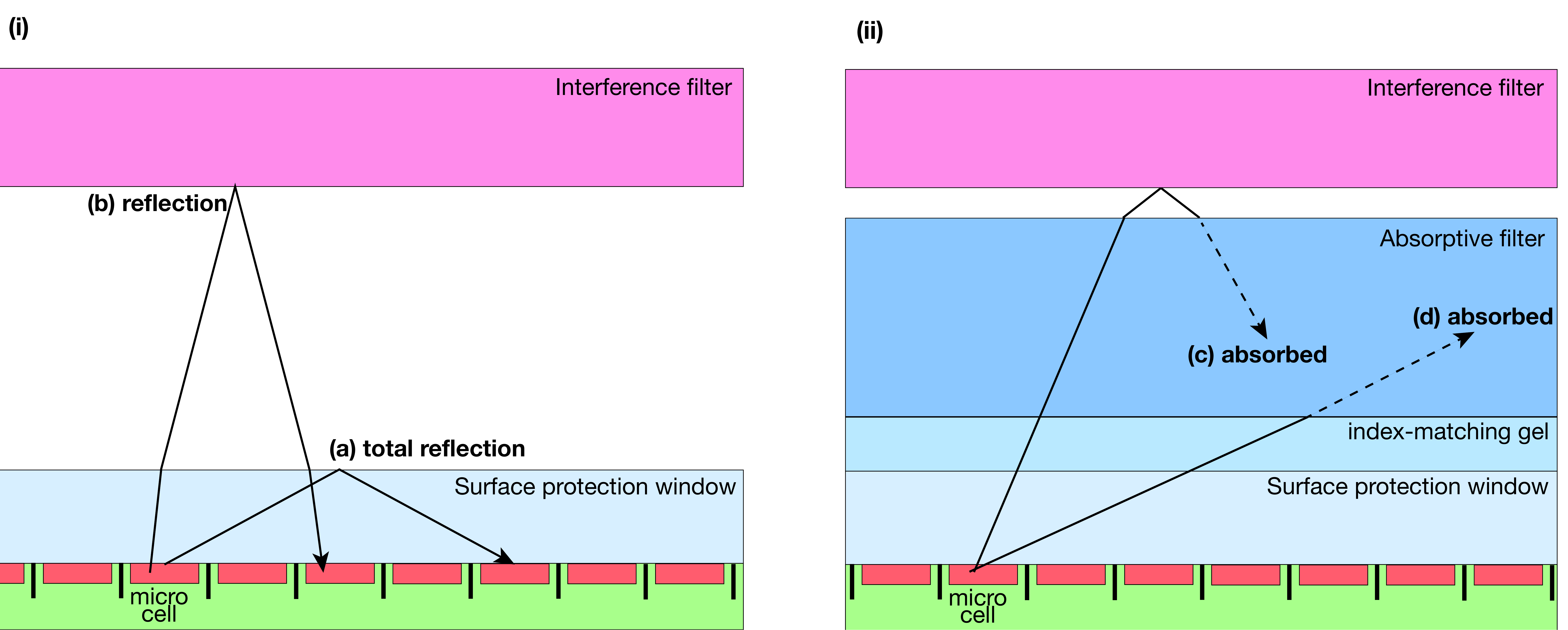}
\caption{(i) schematic of the OCT mechanism in the case of the interference BPF, which corresponds to Fig.~\ref{fig:bpfeffect}(b). (ii) schematic of the OCT suppression mechanism due to the absorptive filter, which corresponds to Fig.~\ref{fig:bpfeffect}(c).}
\label{fig:mechanism}
\end{figure}

This phenomenon can be explained by considering the reflection at the interference BPF as shown in Fig.~\ref{fig:mechanism}(i).
In the figure, one of the micro cells is fired and emits secondary photons. The photon labeled (a) is the original OCT mechanism just as in the case of no BPF; the secondary photon goes back to another micro cell due to the reflection inside the surface protection window \cite{Asano2018} whose thickness is $\sim$0.1\,mm. 
The photon labeled (b) is the additional path due to the interference BPF. The secondary photons are reflected by the BPF and go back to SiPM surface again. The distance between the SiPM surface and the BPF was $\sim$1\,mm, so the secondary photons tend to spread across the whole surface of the SiPM array.
Similar behavior has been reported elsewhere \cite{Mazzillo2017b}; they observed that the dark current was increased when they embedded a green BPF onto the SiPM. Our measurement confirmed that it was surely due to the OCT increase. In addition, this effect is more serious in the case of large area SiPMs or SiPM arrays.

We placed an absorptive BPF between the SiPM and the interference BPF as shown in Fig.~\ref{fig:bpfeffect}(c) to mitigate the OCT increase. We used 1-mm-thick SCHOTT BG40 colored glass whose transmission peak is 500\,nm. We glued the absorptive BPF onto the SiPM surface with index-matching gel (Thorlabs G608N3) whose refractive index is 1.47 at 512\,nm.
The observed OCT probabilities were evidently suppressed. In addition, we found that not only the inter-channel OCT (which was increased due to the interference BPF) but also the intra-channel OCT was well suppressed by a factor of 5 from the original probability without any filter.

The secondary photon paths labeled (c) and (d) in Fig.~\ref{fig:mechanism}(ii) show the mechanism of the suppression. The photon labeled (c) shows the cancelation of the OCT due to reflection at the interference BPF. The photon labeled (d) shows the reduction mechanism of the intra-channel OCT: the secondary photon, which would otherwise be reflected at the protection window surface, is not reflected and enters the absorptive BPF where it is absorbed.
A similar phenomenon has been reported by using NIR LPFs \cite{Mazzillo2018a} for dark current reduction. Our study demonstrated that the OCT reduction can be used in the visible region for the first time. For our original purpose in the ACME experiment where we detect 512\,nm fluorescence induced by a 703\,nm laser, this technique is quite useful to improve SNR: while the photon number is reduced by 3\% after adding the absorptive filter, the total OCT probability is reduced from 24.7\% to 4.3\%. Therefore, the improvement of the photon number resolution is equivalent to a 1.16-fold increase in the photon number.
Because the photon detection efficiency of SiPMs tends to have a peak in the visible region, this technique may be useful for various measurements.

\subsection{OCT measurement with absorptive filters\label{sec:absorptivefilters}}
\begin{figure}[h!]
\centering
\includegraphics[width=13cm, bb=0 0 955 398]{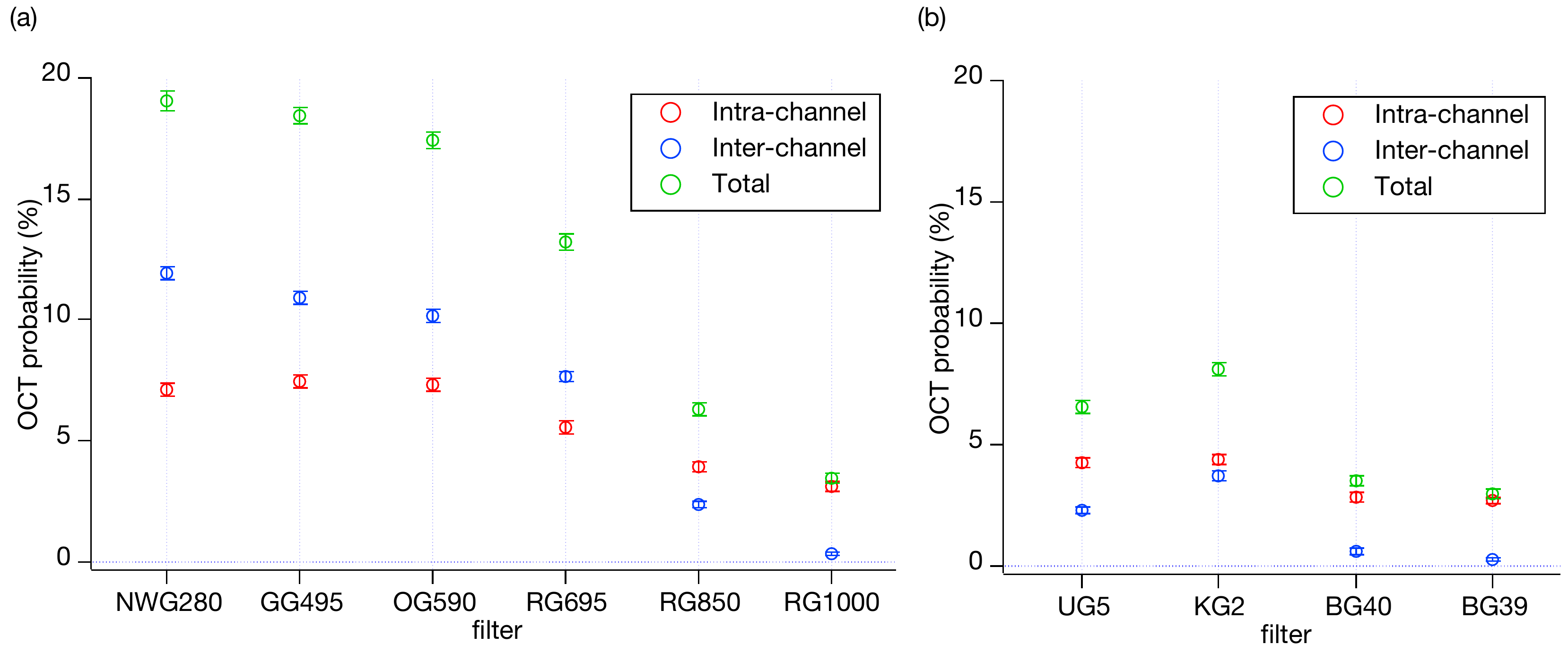}
\caption{(a) OCT probabilities with various absorptive LPFs. The filters are listed from short cut-off to long cut-off filter. (b) OCT probabilities with various absorptive BPFs. The red circles are the intra-channel OCT probabilities of the reference channel. The blue circles show the sum of inter-channel OCT probabilities of the remaining 15 channels. The green circles are the sum of intra-channel and inter-channel OCT probabilities. The error bars represent the statistical uncertainty.}
\label{fig:filterdata}
\end{figure}

The suppression mechanism shown as path (d) in Fig.~\ref{fig:mechanism}(ii) is expected to be quite general.
To generalize this technique, we checked the OCT suppression with absorptive-type LPFs with various cut-on wavelengths. As LPFs, we used SCHOTT NWG280, OG495, OG590, RG695, RG850, and RG1000 ranging from short cut-on to long cut-on wavelengths. All filters were 1-mm-thick plates.
Figure~\ref{fig:filterdata}(a) shows the OCT probabilities for each LPF. The error bars represent statistical uncertainty for one measurement. The reproducibility was $\sim$1\%, probably due to gluing quality.
We also tested several absorptive BPFs: SCHOTT UG5 (80\%), KG2 (89\%), BG40 (12\%), and BG39 (10\%). The numbers in the parentheses represent internal transmittance for a 1\,mm thick filter using 700\,nm photons as a reference. All filters were 1-mm-thick as well. The results are plotted in Fig.~\ref{fig:filterdata}(b). The tendency of the remaining OCT probabilities qualitatively agree with the $\sim$700\,nm transparency.

\section{Discussion\label{sec:discussion}}
\begin{figure}[h]
\centering
\includegraphics[width=11cm, bb=0 0 768 423, clip]{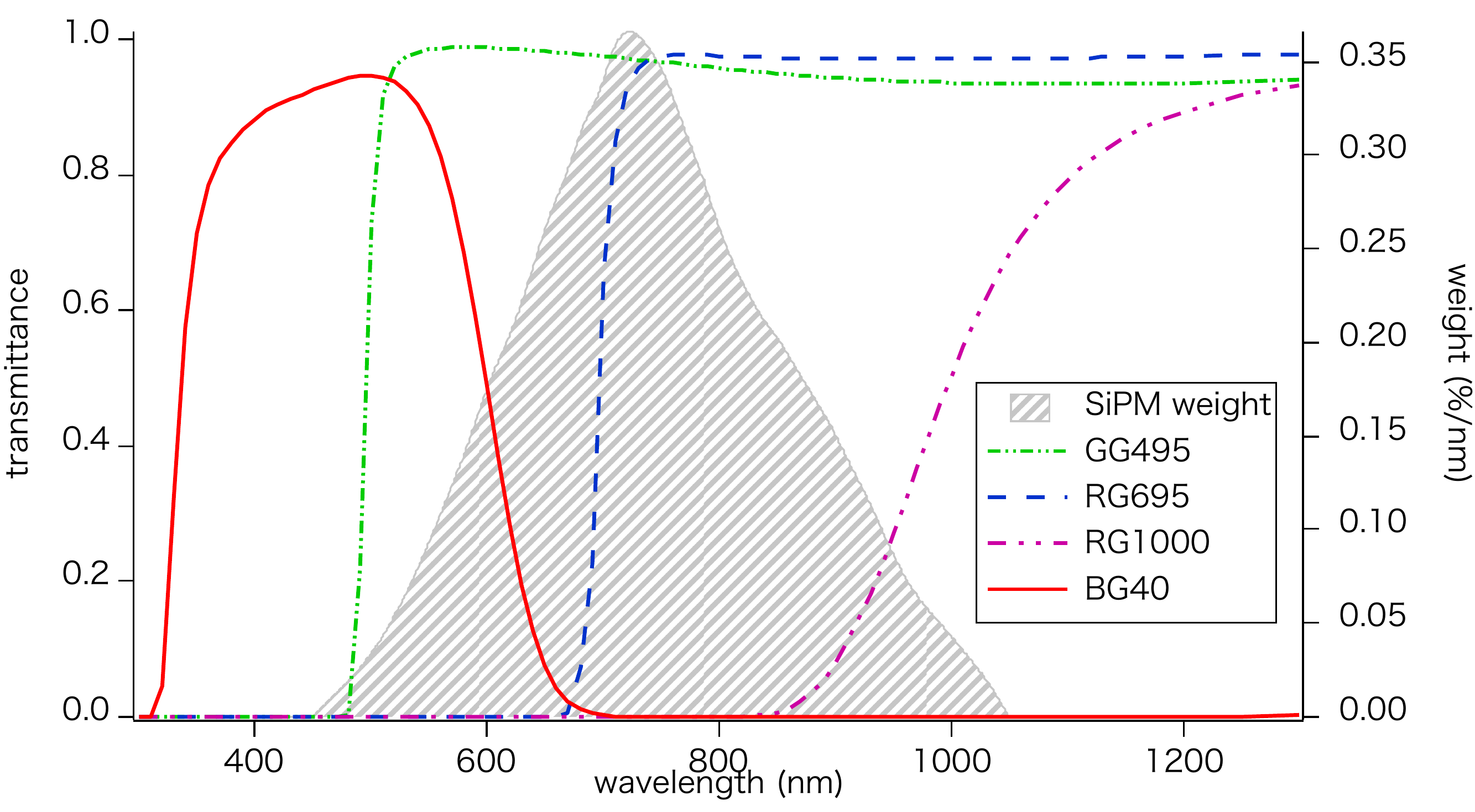}
\caption{Wavelength dependences relevant to the OCT measurement. The hatched region represents the SiPM weight which is the product of the emission spectrum of the secondary photons and the detection efficiency spectrum of the SiPM. The four lines are the internal transmittance curves of 3-mm-thick plates which are quoted from the data sheets.}
\label{fig:spectra}
\end{figure}

These measurements evidently suggest that glueing absorptive filters reduces the OCT probability of both intra-channel and inter-channel mechanisms. Especially, the LPF data set suggests that the wavelength range between 600\,nm to 1000\,nm predominantly contributes to this phenomenon.
This is reasonable because the secondary photons causing the OCT tend to be NIR wavelength  \cite{Mirzoyan2009}. Also, since the SiPM PDE spectrum has a peak in the visible region, the important wavelength region can be estimated.
Figure~\ref{fig:spectra} visually indicates this wavelength dependence. 
The hatched region is the ``weight'' which contributes to the OCT. The weight is a product of two spectra: the secondary photon spectrum quoted from the reference \cite{Mirzoyan2009} and the photon detection efficiency spectrum quoted from the manufacturer's data sheets. It has a peak at $\sim$720\,nm with FWHM of $\sim$250\,nm; that is consistent with the measurement.
We also plot several transmittance curves of the filters used in this study as reference to make the experimental data shown in Fig.~\ref{fig:filterdata} more understandable. The curves are quoted from the manufacturer's data sheets.
These quoted spectra do not perfectly reproduce the present measurements, but these are sufficient to discuss the general trend.

Note that the OCT probabilities with NWG280 and GG495, which are supposed to do nothing, were changed from those without filters: the intra-channel OCT was decreasing while inter-channel OCT was increasing. This was due to the filter thickness which expanded the effective path length of secondary photons. The thickness where secondary photons can propagate was increased from $\sim$0.1\,mm to $\sim$1.1\,mm; thus the secondary photons were able to spread roughly 10 times wider and tended to reach neighboring channels rather than the intra-channel. Such behavior has been reported previously \cite{Nakamura2019}. 
In the case of the RG1000, the inter-channel OCT was almost zero while the intra-channel OCT remained $\sim$3\%. This suggests the remaining OCT occurs inside the silicon plane without entering the surface protection window, through mechanisms such as backside reflection or tunneling through trenches between micro cells. 
We also found that, as expected, neither AP nor DCT probabilities were affected because these phenomena occur inside the silicon plane and are not relevant to the surface protection window.

\section{Conclusion\label{sec:conclusion}}
 We propose a method of optical crosstalk suppression in SiPM systems based on optical filters.
We demonstrate that gluing an absorptive visible bandpass filter onto the SiPM surface can substantially reduce the OCT probability.
Data from a series of measurements using longpass filters and bandpass filters suggest that the absorption of NIR photons is important to  suppress OCT. 
In contrast, interference BPFs increased the OCT probability.
This phenomenon is due to the reflection at the filter surface and it is more serious for large area SiPMs or SiPM arrays.

Optical filters are widely used with photo detectors to improve the SNR; however, this work suggests that it should be carefully designed in a SiPM case due to the OCT. Interference BPFs are often adopted to maximize SNR, but these can increase the OCT probability. 
The proposed technique combines the two, taking full advantage of both the superior background suppression of interference filters and the ability of absorption filters to reduce OCT.

\section*{Appendix A Ray-tracing simulation}

The results shown in Fig.~\ref{fig:filterdata} and the mechanism shown as path (d) in Fig.~\ref{fig:mechanism} were also studied by a simple ray-tracing Monte Carlo simulation.
We built a simulator based on libraries provided by ROOT \cite{Brun1997} and simulated secondary photon propagation inside the protection window and filters.

We assumed all 6400 micro cells in the reference channel are identical and emit secondary photons isotropically and uniformly into the protection window. To simplify the simulation, simple Fresnel reflection at the media boundaries was implemented without any scattering or diffraction. We ignored fine structures of the SiPM such as metal lines and channel boundaries. The reflection at the boundaries of the index matching gel was also ignored because the reflectance is small enough.
The transmittance of the protection window was assumed to be 100\% (fully transparent for any wavelength) and the transmittance of each absorptive filter was quoted from the data sheet. 
We implemented the OCT process to the second order in which the fired cell triggered by the first micro cell successively emit secondary photons again and would trigger the OCT in different channels with the same probability. Because the OCT probability is roughly $\sim$10\%, the second order effect is only $\sim$1\%.

Regarding the intra-channel OCT, there are several paths that secondary photons can reach neighboring micro cells without propagating through the protection window or filter materials. These are not simulated in our study; thus they appear as an offset. On the other hand, inter-channel OCT is produced by the secondary photons propagating through the protection window or filter materials, so they should be included in the simulation. Based on the measurement result, we added a 3\% offset for the intra-channel OCT while doing nothing for the inter-channel OCT.

\begin{figure}[h!]
\centering
\includegraphics[width=13cm, bb=0 0 935 527]{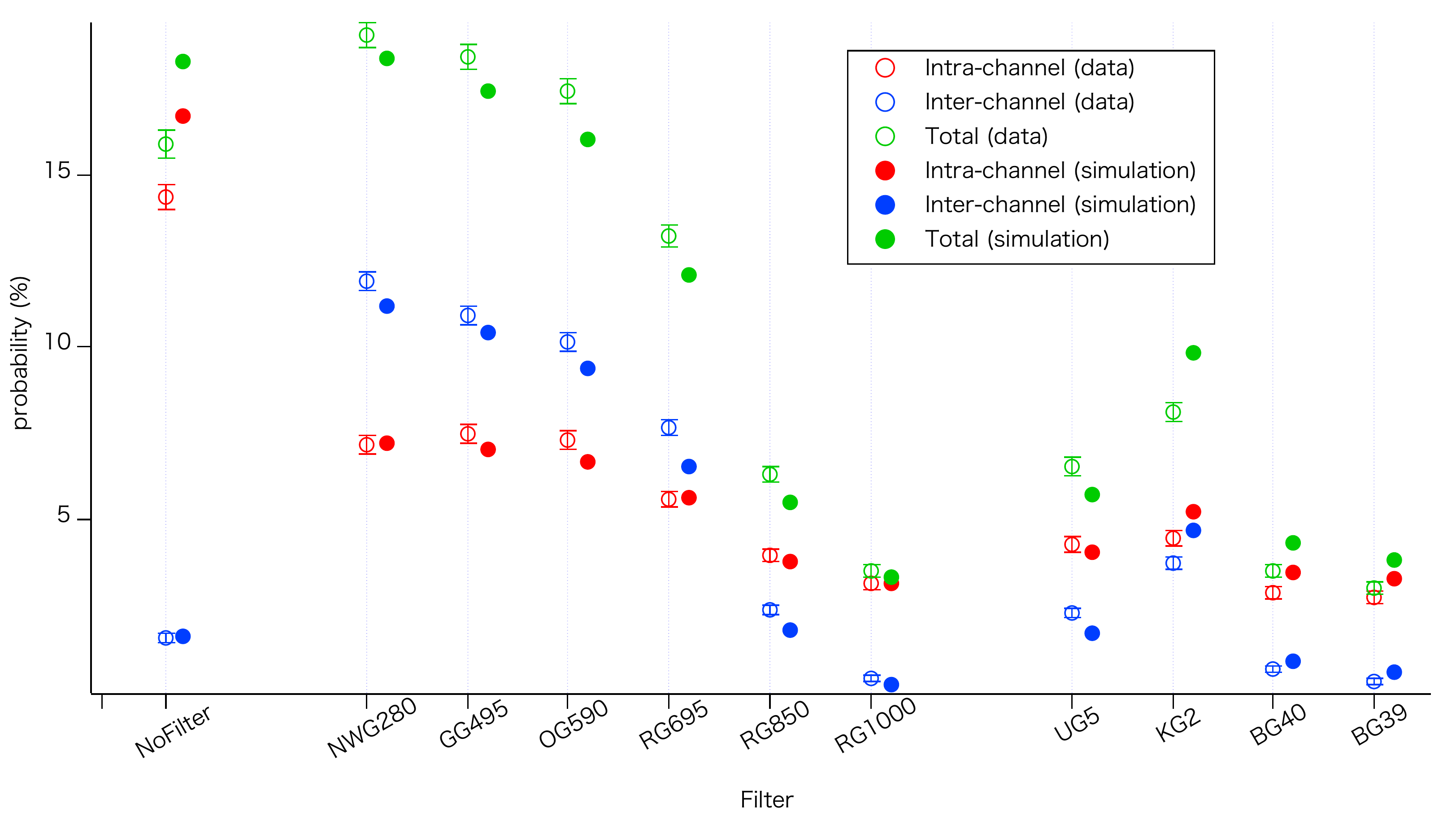}
\caption{Comparison between the experimental data and simulation. The experimental data is the same as that in Fig.~\ref{fig:filterdata}.}
\label{fig:simcomparison}
\end{figure}

Figure~\ref{fig:simcomparison} shows the comparison between the measurement and the simulation. The simulation shows good agreement with the measurement for both the LPFs and the BPFs, even though we did not do any fine tuning for the wavelength dependence. 
The simulation supports the reduction mechanism quantitatively. 
During the simulator construction, we found that the ratio between the intra-channel and inter-channel OCT depends on the geometric configuration, especially the thickness of the protection window, gel, and filters, and the angular distribution of the secondary photon emission. Since the protection window and gel thickness were ambiguous and not uniform while the filter thickness was well controlled, we roughly adjusted the thickness so as to reproduce the intra-channel and inter-channel OCT ratio with filters rather than the no filter configuration. The resultant thickness of the protection window and gel were 150\,$\mu$m and 50\,$\mu$m, respectively. 
Also, we empirically applied a cut in the angular distribution as $\theta < 1.3$\,rad, meaning very shallow paths were ignored. As a result, the reproducibility of the simulation in the no filter configuration was slightly worse than the configurations with filters. This may be due to the ambiguity of the protection window. As shown in Fig.~\ref{fig:bpfeffect}(a), for example, the inter-channel OCT probabilities of the adjacent channels were not uniform while the simulation suggests a uniform distribution. It could be due to non-flatness of the protection window that was not considered in the simulation.
Overall, the simulation is consistent with the experimental data. In particular, it explains the results of the LPF configuration well.

\section*{Funding}
This work was supported by JSPS KAKENHI Grant Number JP20KK0068 and the RECTOR program in Okayama University.

\section*{Acknowledgments}
We would like to thank T. Kobayakawa for the basic characterization of the SiPM module, and C. Panda and Z. Lasner for initial considerations of using SiPMs. We also appreciate D. Lascar, Z. Han, P. Hu, S. Liu and B. Hao for helpful discussions.

\section*{Disclosures}
The authors declare no conflicts of interest.

\bibliography{OCT}

\end{document}